\documentclass[twocolumn]{aastex62}

\usepackage{graphicx,amsmath}
\usepackage{amssymb}
\usepackage{subfigure}
\usepackage{color}
\usepackage{soul}
\usepackage{CJK}
\hypersetup{linkcolor=blue, citecolor=blue}

\newcommand{\hi}{H\thinspace{\sc i}}
\newcommand{\hii}{H\thinspace{\sc ii}}
\newcommand{\oiii}{[O\thinspace{\sc iii}]}
\newcommand{\nii}{[N\thinspace{\sc ii}]}

\newcommand{\htwo}{H$_2$}
\newcommand{\ha}{H$\alpha$}
\newcommand{\hb}{H$\beta$}

\newcommand{\Msol}{\hbox{\thinspace $M_{\odot}$}}

\defcitealias{Zhang2019}{Z19}	

\graphicspath{{./}{Figures_v2/}}

\received{August 11, 2020}
\published{April 15, 2021}

\shorttitle{The death of massive disk galaxies}
\shortauthors{Zhang et al.}

\begin{document}

\title{\bf \large Mass and Environment as Drivers of Galaxy Evolution. IV. On the Quenching of Massive Central Disk Galaxies in The Local Universe}

\correspondingauthor{Yingjie Peng}
\email{yjpeng@pku.edu.cn}

\author{Chengpeng Zhang}
\affil{Kavli Institute for Astronomy and Astrophysics, Peking University, 5 Yiheyuan Road, Beijing 100871, P. R. China}
\affil{Department of Astronomy, School of Physics, Peking University, 5 Yiheyuan Road, Beijing 100871, P. R. China}

\author{Yingjie Peng}
\affiliation{Kavli Institute for Astronomy and Astrophysics, Peking University, 5 Yiheyuan Road, Beijing 100871, P. R. China}

\author{Luis C. Ho}
\affil{Kavli Institute for Astronomy and Astrophysics, Peking University, 5 Yiheyuan Road, Beijing 100871, P. R. China}
\affil{Department of Astronomy, School of Physics, Peking University, 5 Yiheyuan Road, Beijing 100871, P. R. China}

\author{Roberto Maiolino}
\affiliation{Cavendish Laboratory, University of Cambridge, 19 J. J. Thomson Avenue, Cambridge CB3 0HE, UK}
\affiliation{Kavli Institute for Cosmology, University of Cambridge, Madingley Road, Cambridge CB3 0HA, UK}
\affiliation{Department of Physics and Astronomy, University College London, Gower Street, London WC1E 6BT, UK}

\author{Alvio Renzini}
\affiliation{INAF - Osservatorio Astronomico di Padova, Vicolo dell'Osservatorio 5, I-35122 Padova, Italy}

\author{Filippo Mannucci}
\affiliation{Istituto Nazionale di Astrofisica, Osservatorio Astrofisico di Arcetri, Largo Enrico Fermi 5, I-50125 Firenze, Italy}

\author{Avishai Dekel}
\affiliation{Racah Institute of Physics, The Hebrew University, Jerusalem 91904, Israel}

\author{Qi Guo}
\affiliation{Key Laboratory for Computational Astrophysics, National Astronomical Observatories, \\Chinese Academy of Sciences, Beijing 100012, P. R. China}
\affiliation{School of Astronomy and Space Science, University of Chinese Academy of Sciences, Beijing 100049, P. R. China}

\author{Di Li}
\affiliation{CAS Key Laboratory of FAST, National Astronomical Observatories, Chinese Academy of Sciences, Beijing 100012, P. R. China}
\affiliation{School of Astronomy and Space Science, University of Chinese Academy of Sciences, Beijing 100049, P. R. China}

\author{Feng Yuan}
\affiliation{Key Laboratory for Research in Galaxies and Cosmology, Shanghai Astronomical Observatory, Chinese Academy of Sciences, 80 Nandan Road, Shanghai 200030, P. R. China}

\author{Simon J. Lilly}
\affiliation{Department of Physics, ETH Zurich, Wolfgang-Pauli-Strasse 27, CH-8093 Zurich, Switzerland}

\author{Jing Dou}
\affil{Kavli Institute for Astronomy and Astrophysics, Peking University, 5 Yiheyuan Road, Beijing 100871, P. R. China}
\affil{Department of Astronomy, School of Physics, Peking University, 5 Yiheyuan Road, Beijing 100871, P. R. China}

\author{Kexin Guo}
\affil{Kavli Institute for Astronomy and Astrophysics, Peking University, 5 Yiheyuan Road, Beijing 100871, P. R. China}
\affil{International Centre for Radio Astronomy Research, University of Western Australia, Crawley, WA 6009, Australia}

\author{Zhongyi Man}
\affil{Kavli Institute for Astronomy and Astrophysics, Peking University, 5 Yiheyuan Road, Beijing 100871, P. R. China}
\affil{Department of Astronomy, School of Physics, Peking University, 5 Yiheyuan Road, Beijing 100871, P. R. China}

\author{Qiong Li}
\affil{Kavli Institute for Astronomy and Astrophysics, Peking University, 5 Yiheyuan Road, Beijing 100871, P. R. China}
\affil{Department of Astronomy, School of Physics, Peking University, 5 Yiheyuan Road, Beijing 100871, P. R. China}

\author{Jingjing Shi}
\affil{Kavli Institute for Astronomy and Astrophysics, Peking University, 5 Yiheyuan Road, Beijing 100871, P. R. China}
 \affil{Kavli IPMU (WPI), UTIAS, The University of Tokyo, Kashiwa, Chiba 277-8583, Japan}

\begin{abstract}

\noindent The phenomenological study of evolving galaxy populations has shown that star forming galaxies can be quenched by two distinct processes: mass quenching and environment quenching \citep{Peng2010}. To explore the mass quenching process in local galaxies, we study the massive central disk galaxies with stellar mass above the Schechter characteristic mass. In \citet{Zhang2019}, we showed that during the quenching of the massive central disk galaxies as their star formation rate (SFR) decreases, their molecular gas mass and star formation efficiency drop rapidly, but their \hi~gas mass remains surprisingly constant. To identify the underlying physical mechanisms, in this work we analyze the change during quenching of various structure parameters, bar frequency, and active galactic nucleus (AGN) activity. We find three closely related facts. On average, as SFR decreases in these galaxies: (1) they become progressively more compact, indicated by their significantly increasing concentration index, bulge-to-total mass ratio, and central velocity dispersion, which are mainly driven by the growth and compaction of their bulge component; (2) the frequency of barred galaxies increases dramatically, and at a given concentration index the barred galaxies have a significantly higher quiescent fraction than unbarred galaxies, implying that the galactic bar may play an important role in mass quenching; and (3) the ``AGN'' frequency increases dramatically from 10\% on the main sequence to almost 100\% for the most quiescent galaxies, which is mainly driven by the sharp increase of LINERs. These observational results lead to a self-consistent picture of how mass quenching operates. 

\end{abstract}


\section{Introduction} \label{sec:intro}
Local galaxies can be broadly divided into star-forming and quiescent (or passive) galaxies \citep[e.g.][]{Blanton2003a,Kauffmann2003,Bell2004,Brinchmann2004,Baldry2006self,Peng2010}.  Quiescent galaxies were actively forming stars at higher redshifts \citep{Daddi2005self,Santini2009} and their star formation was quenched later by one or more mechanisms. Identifying the physical mechanism responsible for star formation quenching becomes one of the most debated open questions. It has now been well established that the star formation activities in the local galaxies critically depend on their stellar mass \citep[e.g.,][]{Kauffmann2003self,Baldry2006self,Peng2010} and environment \citep[e.g.,][]{Bosch2008self,Peng2012,Woo2013Mself,Bluck2016}. In the phenomenological study of the evolving galaxy population of \citet{Peng2010}, the differential effects of stellar mass and environment on the fraction of galaxies that are on the red sequence, $f_{\rm red}$, are found to be completely separable. This hence suggested that two distinct processes are operating, one called ``mass quenching,'' which is independent of environment, and the other called ``environment quenching,'' which is independent of stellar mass.

Both mass quenching and environment quenching have been studied extensively in the past decades. The physical mechanisms proposed to be responsible for environment quenching include strangulation \citep{Larson1980,Balogh2000,Balogh2000b,Peng2015}, ram pressure stripping \citep{Gunn1972,Abadi1999,Quilis2000}, tidal stripping and harassment \citep{Farouki1981,Moore1996}, major merger \citep{Mihos1996self,Hopkins2008self} and halo quenching \citep{Dekel2006self}. On the other hand, the physical mechanisms proposed to be responsible for mass quenching include active galactic nucleus (AGN) feedback \citep{Croton2006,Fabian2012,Harrison2017a}, morphological quenching \citep{Martig2009}, gravitational quenching \citep{Genzel2014}, dynamical quenching in galaxy spheroids \citep{Gensior2020}, bar quenching \citep{Gavazzi2015b,Khoperskov2018self} and angular momentum quenching \citep{Peng2020,Renzini2020}.  

It has been long debated which is the main driver of mass quenching among all the possible mechanisms. For the above-proposed mechanisms, they often come with both positive and negative evidence. For example, \citet{Maiolino2012self} find quasar-driven massive outflow in the early universe can clean the gas in the host galaxy and quench star formation rapidly. \citet{Terrazas2017self} find that the specific star formation rate (SFR) decreases with increasing black hole mass in local galaxies. Meanwhile, many other observations find no evidence of AGN feedback quenching. For instance, \citet{Shangguan2018self,Shangguan2020} find that most low-redshift quasar hosts have similar gas content compared to those of massive star-forming galaxies. The \hi~gas reservoirs of local AGN host galaxies are also found to be normal compared to non-AGN hosts \citep{Ho2008HI,Ellison2019}. Positive AGN feedback that triggers star formation has also been observed in AGN hosts \citep{Cresci2015self,Maiolino2017self}. As is well known, the cosmic evolution of SFR density and AGN activity parallel each other, both rising from high redshifts, peaking at $z\sim2$, and then decreasing to the present, in the local universe \citep{Madau2014self}. This is not what one would naively expect if AGNs could kill star formation. Rather, it suggests that star formation and nuclear activity are both fueled by the available gas, with no direct responsibility of AGNs in quenching star formation.

As the fuel of star formation, the cold gas content in galaxies will provide direct observational evidence of how quenching may happen. In order to study the mass quenching mechanisms, we analyzed the atomic and molecular gas content for massive central galaxies in \citet[hereafter \citetalias{Zhang2019}]{Zhang2019} by using the sample of ALFALFA, GASS, and COLD GASS surveys. Our results reveal that disk galaxies with SFR well below the main sequence surprisingly have the same large atomic hydrogen (\hi) gas reservoir as that of star-forming galaxies, which is unexpected because the galaxies undergoing star formation quenching were thought to have less cold gas \citep[e.g.][]{Fabello2011self,Huang2012,Brown2015,Saintonge2016self,Catinella2018self,
Tacconi2018self}. In contrast with \hi~gas, \citetalias{Zhang2019} further showed that the molecular gas mass and star formation efficiency of central disk galaxies in the process of being quenched are significantly lower than those of star-forming galaxies, which is the direct causation of their low-level star formation activities. The lower molecular gas masses of massive disk galaxies are also revealed by some recent observations \citep{Brownson2020self,Luo2020self}. Our findings in \citetalias{Zhang2019} clearly show how quenching proceeds in these massive central galaxies. However, the physical mechanisms that reduce their molecular gas amount and star formation efficiency and therefore quench the star formation remain unclear.


Since the extended \hi~gas is very sensitive to environmental effects \citep{Giovanelli1985,Catinella2013}, the unchanged \hi~gas reservoir in the quenching process of massive central disk galaxies (as shown in \citetalias{Zhang2019}) corroborates that the star formation of these galaxies is halted by internal processes. One possible mechanism is the buildup of the bulge component in galaxies, which can stabilize the gas against fragmentation to bound clumps and reduce the star formation efficiency, through morphological quenching \citep{Martig2009,Gensior2020} or gravitational quenching \citep{Genzel2014}. Recent works showed that the quenched fraction of galaxies strongly correlates with the presence of the central spheroid structure, as indicated by the measurable structural parameters, such as bulge-to-total ratio (B/T), S\'ersic index ($n_{\rm s}$), concentration index ($R_{90}$/$R_{50}$), and central velocity dispersion ($\sigma_*$) \citep[e.g.,][]{Cheung2012self, Wake2012self, Bluck2014self}. 

The galactic bar is another key driver of the internal secular evolution of disk galaxies. A strong bar is able to drive gas inflows toward galactic center \citep{Athanassoula1992,Sheth2005} and produce a gas-deficient region on the scale of several kiloparsecs \citep{Gavazzi2015b,Spinoso2017}. Some case studies or statistics based on small samples reveal clear signatures of suppressed star formation in the bar region \citep{James2016,George2019M95,Krishnarao2020,Newnham2020}.
Based on the visual selection of barred galaxies, \citet{Masters2010self} and \citet{Guo2020self} showed that massive spiral galaxies with red optical color have much higher bar fraction compared to blue spiral galaxies.

Another internal quenching mechanism that is often proposed is the feedback from AGNs \citep{Croton2006,Fabian2012,Maiolino2012self,Harrison2017a}. AGN feedback has been widely implemented as a key recipe in most semianalytic models \citep[e.g.][for a review]{Baugh2006} and hydrodynamical simulations \citep{Schaye2015,Weinberger2018self, Yuan2018self}. Observationally, the AGN host galaxies are more likely to be found in the transitional galaxies (green valley) in the process of being quenched \citep{Nandra2007,Schawinski2007,Leslie2016,Silverman2019}.

Overall, all of the internal drivers of galaxies mentioned above are found to correlate with the star formation quenching of massive galaxies. However, most previous studies were not aimed at separating the internal quenching mechanism from the external, environmental one. In this paper, we focus on massive central disk galaxies as in \citetalias{Zhang2019}. The detailed reasons of restricting the sample to central disk galaxies will be justified in Section \ref{sec:sample}. We study the change of their internal properties in the quenching process to investigate the possible physical origin of mass quenching, including the effect of massive bulge, bar instability, and AGN feedback. This paper is structured as follows. Section \S2 describes the sample selection and parameter determination. In Section \S3, we present the main results and discussion, including the study of key structural parameters (Section \S3.1), possible AGN feedback (Section \S3.2) and the impact of using different SFR indicators (Section \S3.3). We summarize our results in Section \S4. The \citet{Chabrier2003self} initial mass function (IMF) is used throughout this work. We assume the following cosmological parameters: $\Omega_m=0.3, \Omega_\Lambda=0.7, H_0=70\,\rm {km\,s^{-1}\,Mpc^{-1}}$.


\section{Sample} \label{sec:sample} 

\subsection{Why Massive Central Disk Galaxy?}

As in \citetalias{Zhang2019}, we focus on massive central disk galaxies to study the mass quenching process in this work. The reasons for restricting the sample to central disk galaxies are as follows. First, many ellipticals lying on the passive sequence are quenched at high redshifts \citep[e.g. $z\sim2-3$; ][]{Bower1992self,Daddi2005self,Lucia2006self,Onodera2012,DEugenio2020self}. Including these ellipticals in our analysis will not help us to understand the quenching mechanism in the local universe but introduces complicated progenitor bias \citep{Lilly2016a}. Second, there is a consensus that in the local universe disks are transformed into ellipticals mainly by mergers \citep{Toomre1972,Barnes1992,Naab2003,Bournaud2005a}. There is no evidence that any internal quenching mechanism, such as AGN feedback, is able to transform disks to ellipticals. Third, elliptical galaxies and disk galaxies have very different structural parameters (e.g., $n_{\rm s}$, $R_{90}$/$R_{50}$ and $\sigma_*$). The statistical changing structure in the quenching process found in previous works \citep[e.g.][]{Cheung2012self, Wake2012self, Bluck2014self} is largely driven by the increasing fraction of elliptical galaxies, not the continuously secular evolution of disk galaxies during quenching.
As shown in \citetalias{Zhang2019}, by focusing on massive central disk galaxies at fixed stellar mass range, we found that quenched central disk galaxies or disk galaxies in the process of being quenched surprisingly have the same large \hi~gas reservoir as star-forming ones. This new result manifests the importance of sample selection in the study of quenching mechanisms. The mean \hi~gas amount in these low-SFR galaxies will be significantly less than that of star-forming galaxies if we involve ellipticals or satellites in our analysis.

\subsection{The Sloan Digital Sky Survey} \label{sec:SDSS}

The parent galaxy sample analyzed in this paper is the same Sloan Digital Sky Survey (SDSS) DR7 \citep{Abazajian2009} sample that we constructed in \citet{Peng2010,Peng2012,Peng2015} for similar statistical investigations of star formation and the quenching process. Briefly, it is a magnitude-selected sample of galaxies that have clean photometry and Petrosian SDSS $r$-band magnitudes in the range of 10.0$-$18.0 after correcting for Milky Way extinction. The parent photometric sample contains 1,579,314 objects after removing duplicates, of which 238,474 have reliable spectroscopic redshift measurements in the redshift range of $0.02<z<0.085$. To statistically correct the incompleteness effect, each galaxy is weighted by 1/TSR, where TSR is a spatial target sampling rate, determined using the fraction of objects that have spectra in the parent photometric sample within the minimum SDSS fiber spacing of 55\arcsec~of a given object. In the stellar mass range $10^{10.6}$$-$$10^{11}\Msol$ and redshift range 0.02$-$0.085 concerned in this paper, our sample is complete and no volume correction is needed.

\begin{figure}[t!]
\includegraphics[width=\columnwidth]{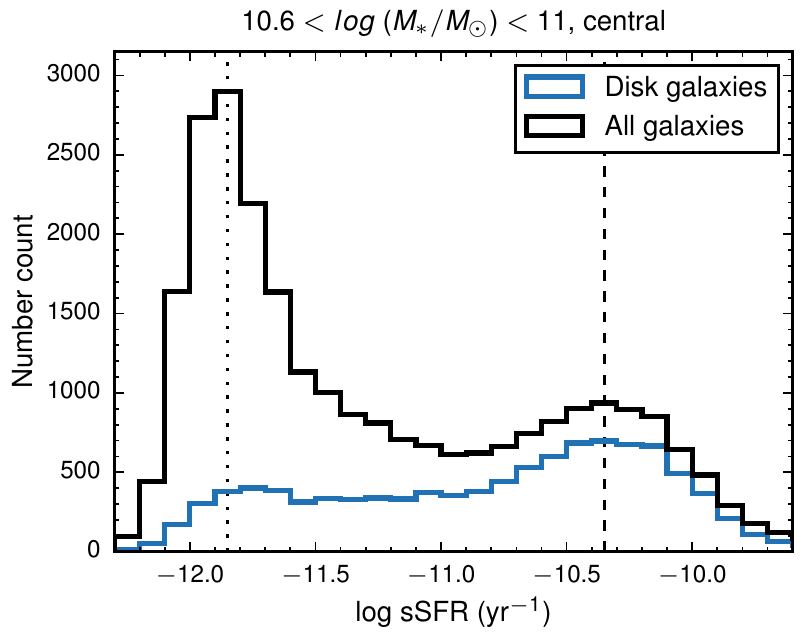}
\caption{The sSFR distribution of central galaxies in the stellar mass range of $10^{10.6}$$-$$10^{11}\Msol$. The distribution of disk galaxies classified by Galaxy Zoo is plotted by the blue histogram. The distribution of all galaxies, including disks, ellipticals, and uncertains, is plotted by the black histogram. The dashed line and dotted line indicate the ridge of the star-forming main sequence and the passive sequence, respectively.}
\label{fig:pdf}
\end{figure}

The stellar masses ($M_*$) used in this paper are determined from the $k$-correction program v4\_1\_4 \citep{Blanton2007} with population synthesis models of \citet{Bruzual2003}. The derived stellar masses are highly consistent with the published stellar masses of \citet{Kauffmann2003} with a small scatter of $\sim$0.1 dex. The central stellar velocity dispersion $\sigma_*$ of each galaxy is derived by the Wisconsin Group \citep{Chen2012,Maraston2013self,Thomas2013self} from the optical rest-frame spectra using a principal component analysis (PCA) method. 

We classify our sample into central galaxies and satellite galaxies using the SDSS DR7 group catalog from \citet{Yang2005self,Yang2007}. To reduce the contamination of the central sample by spurious interlopers into the group, we define central galaxies to be simultaneously both the most massive and the most luminous (in $r$ band) galaxy within a given group. The centrals also include single galaxies that do not have identified companions above the SDSS flux limit. 

\begin{figure*}[ht]
\includegraphics[width=180mm]{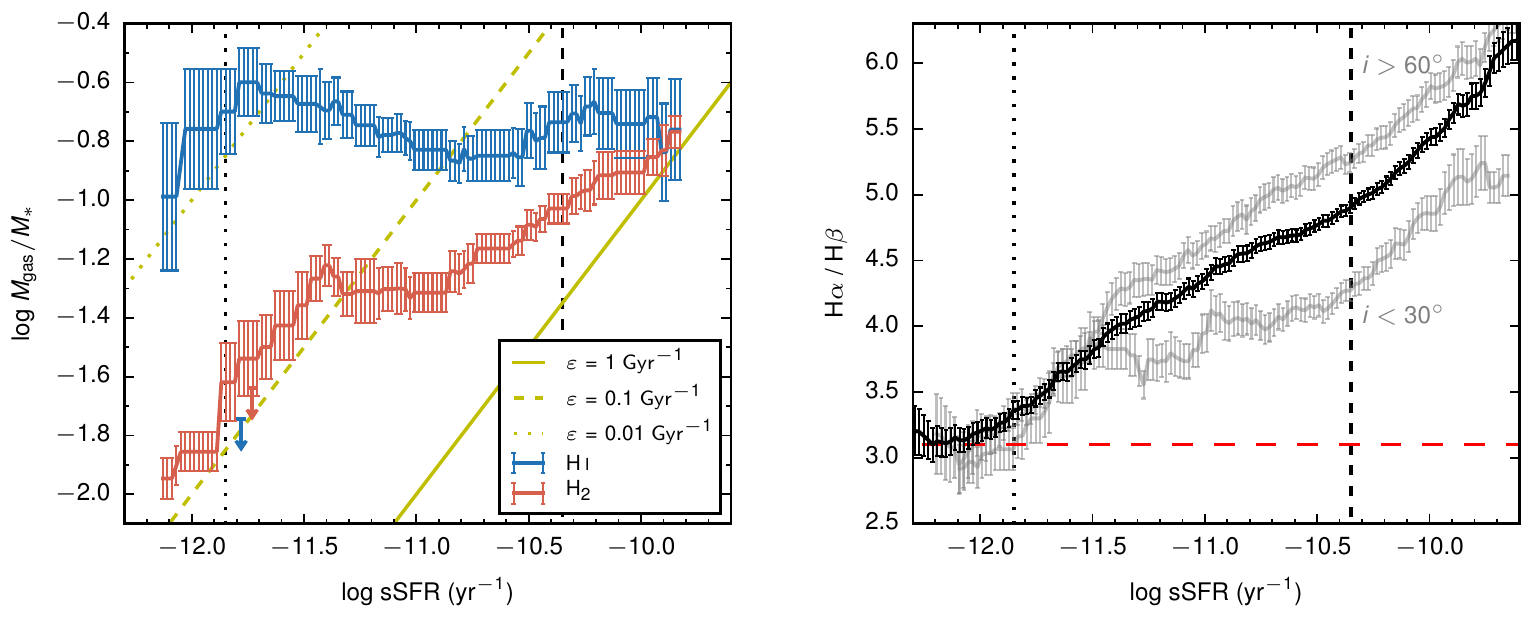}
\caption{Atomic gas, molecular gas, and Balmer decrement for massive central disk galaxies in the stellar mass range of $10^{10.6}$$-$$10^{11}\Msol$. This is a reprise of the main result of \citetalias{Zhang2019} by replacing the x-axis with sSFR to better minimize the effect of the stellar mass dependence of gas mass. \textbf{Left:} The average \hi~mass fraction ($M_{\rm HI}/M_*$, in blue) and \htwo~mass fraction ($M_{\rm H_2}/M_*$, in red) for central disk galaxies in the stellar mass range of $10^{10.6}$$-$$10^{11}\Msol$ in COLD GASS survey. Here we use gas-to-stellar mass ratio instead of using absolute gas mass to further weaken the effect of varying stellar mass in our sample. The blue arrow shows the upper limit of the \hi~mass fraction of the only galaxy without \hi~detection, and the red arrow shows the upper limit of the \htwo~mass fraction of the only galaxy without \htwo~detection. These upper limits are not included to derive the average gas fraction. The yellow lines indicate the constant star formation efficiency for three different values. \textbf{Right:} The average value of Balmer decrement (H$\alpha$/H$\beta$) for all central disk galaxies (black line), for those with $i<30^{\circ}$ (nearly face-on; lower gray line), and for those with $i>60^{\circ}$ (nearly edge-on; upper gray line), within the same stellar mass range in SDSS. The horizontal dashed line indicates the intrinsic value of 3.1 without dust extinction. In each panel, the vertical dashed line and dotted line indicate the ridge of the star-forming main sequence and the passive sequence, respectively. The average values are calculated in the sliding box of 0.5\,dex in sSFR, and the error bars are standard errors with 1$\sigma$ uncertainty.}
\label{fig:gas}
\end{figure*}

\subsection{The Star Formation Rates} \label{sec:SFR} 

The SFRs of our sample are taken from the value-added MPA-JHU SDSS DR7 catalog \citep{Brinchmann2004} and converted to Chabrier IMF by using log SFR (Chabrier) = log SFR (Kroupa) $-0.04$. These SFRs are based on H$\alpha$ emission-line luminosities, corrected for extinction using the H$\alpha$/H$\beta$ ratio. To correct for the aperture effects, the SFRs outside the SDSS 3\arcsec~fiber were obtained by performing the spectral energy distribution (SED) fitting to the {\it ugriz} photometry outside the fiber, using the models and methods described in \citet{Salim2007}. Since the H$\alpha$ emission of AGN and composite galaxies are likely to be contaminated by their nuclear activity, their SFRs are derived based on the strength of
the 4000\,$\mbox{\normalfont\AA}$ break as calibrated with H$\alpha$ for non-AGN, pure star-forming galaxies \citep[see details in][]{Brinchmann2004}. Different SFR estimators may produce different results and, therefore, we repeat all our analysis adopting the widely used SFRs obtained from the SED fitting of UV, optical, and mid-IR
bands \citep[GSWLC-M2 catalog,][]{Salim2016a,Salim2018self}. The results will be presented and discussed in Section \ref{sec:SFR_Salim}.

\subsection{Morphology Classification} \label{sec:disk}

We utilize the morphology classifications from the Galaxy Zoo (GZ) project
\citep{Lintott2011a}, which have been widely used in many previous studies on the disk and spiral galaxies \citep[e.g.,][]{Masters2010self,Hao2019,Guo2020self}. In GZ, the image of each SDSS galaxy was viewed and classified by dozens of volunteers (with an average number of 39) to determine whether the galaxy has a disk or spiral arms. After a careful debiasing process, a morphology flag (``spiral,'' ``elliptical,'' or ``uncertain'') is assigned to each galaxy and is used in this work. Most lenticular or S0 galaxies with smooth and rounded profiles are classified in GZ as ``elliptical'' or ``uncertain.'' Since ``spirals'' in GZ include disk galaxies with or without clear spiral arms, we simply designate all galaxies classified as ``spiral'' in GZ as ``disk'' galaxies. In total, 3\% of these disk galaxies are excluded because their vote fractions of merger (``P\_MG'' in the GZ catalog) are greater than 0.3, which indicates that they are very likely mergers.

Meanwhile, people also propose using structural parameters to define disk galaxies, for instance, \citet{Cortese2020} used B/T to select disk galaxies and found different results compared to those of using visual morphology classification. However, as shown in \citetalias{Zhang2019}, the disk galaxies selected from GZ have an average \hi~detection ratio of $>$90\%, while the \hi~detection ratio of elliptical galaxies in GZ is only $<$20\%. It is not possible to reproduce this result using any other quantitative structural parameters (such as B/T, $R_{90}/R_{50}$, S\'ersic index, and $\sigma_*$), since most of the massive disk galaxies with low sSFR have a massive bulge and their structural parameters are comparable to those of elliptical galaxies, as we will show in Section 3.1 of this paper. Therefore, as an observational effect, the visual morphology classification is more effective to distinguish \hi-rich disk galaxies and \hi-poor galaxies. Why it is important to focus on \hi-rich galaxies? As discussed in \citetalias{Zhang2019}, the symmetric characteristic double-horn \hi~profiles of these galaxies suggest that they have regularly rotating \hi~disks, and the radii of these \hi~disks are about 30\,kpc according to the \hi~mass$-$size relation \citep{Broeils1997self,Wang2016self}. If the extended \hi~gas reservoir decreases in the quenching process, it indicates that some external quenching mechanism is acting in suppressing star formation, and therefore we cannot purely isolate the mass quenching process. Thus, we follow \citetalias{Zhang2019} to select disk galaxies using the visual morphology classification from GZ.


\begin{figure*}[t!]
\includegraphics[width=180mm]{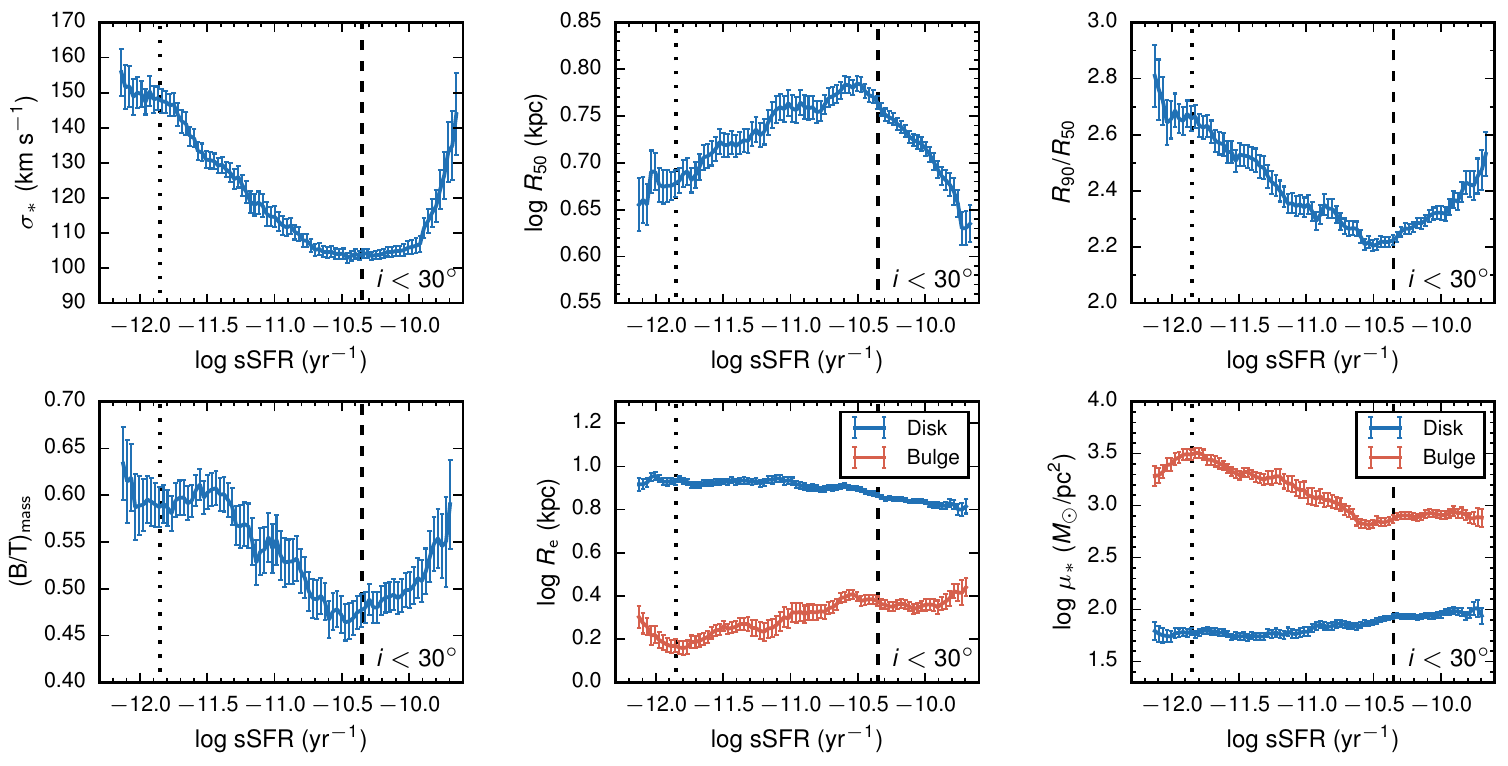}
\caption{Change of key structural parameters as a function of sSFR for central disk galaxies within the stellar mass range of $10^{10.6}$$-$$10^{11}\Msol$. The top three panels show the mean stellar velocity dispersion ($\sigma_*$), half-light radii ($R_{50}$) in $r$ band, concentration index ($R_{90}$/$R_{50}$) in $r$ band. The bottom three panels show the mean bulge-to-total mass ratio (B/T), effective radius ($R_{\rm e}$) in $r$ band of the bulge component and disk component, and surface mass density ($\mu_*$ = $M_*/(2\pi R_{\rm e}^2)$) of the bulge component and disk component. Only about face-on galaxies ($i<30^{\circ}$) are selected. In each panel, the dashed line and dotted line indicate the ridge of the star-forming main sequence and the passive sequence, respectively. The average values are calculated in the sliding box of 0.3\,dex in sSFR, and the error bars are standard errors with 1$\sigma$ uncertainty.}
\label{fig:str}
\end{figure*}

\vspace{0.2cm}
In total, there are 25,800 central galaxies in the stellar mass bin $10^{10.6}$$-$$10^{11}\Msol$ in our sample. The bimodality distribution of the sSFR ($\equiv$ SFR/$M_*$) of these galaxies is shown as the black histogram in Figure \ref{fig:pdf}. The two peaks of the distribution labeled by the dashed line and dotted line in Figure \ref{fig:pdf} indicate the ridges of the star-forming main sequence and passive sequence, respectively, in this stellar mass range. It should be noted that, as discussed in \citet{Renzini2015}, the existence of the quenched peak (i.e., a passive sequence) is due to the large number of galaxies with just SFR upper limits. According to the morphology classification of GZ, 10,083 disk galaxies were selected for further analysis in this paper, as shown by the blue histogram in Figure \ref{fig:pdf}.

\begin{figure}[t]
    \begin{center}
       \includegraphics[width=\columnwidth]{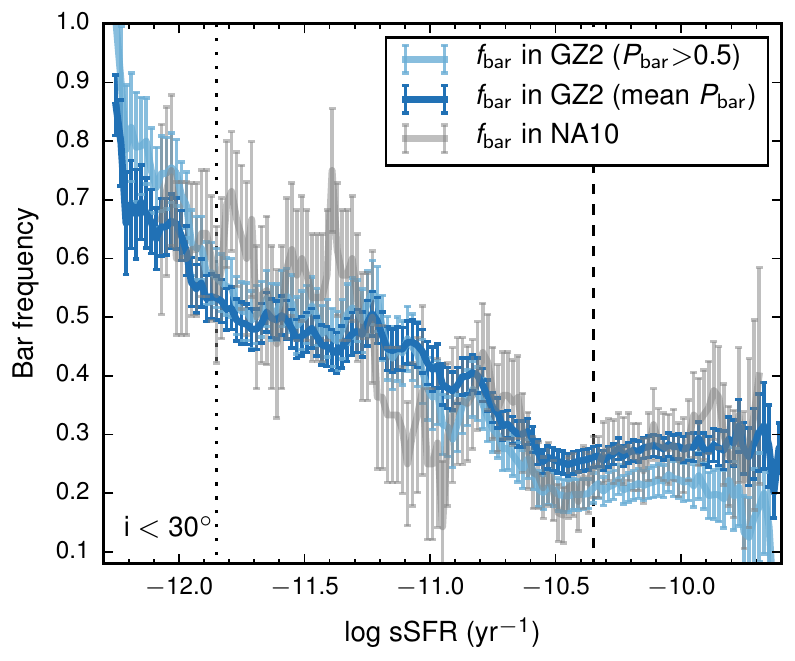}
    \end{center}
\caption{Bar frequency for massive central disk galaxies as a function of sSFR within the stellar mass range of $10^{10.6}$$-$$10^{11}\Msol$. Barred galaxies are selected by two independent works both based on visual classification of SDSS galaxies: Galaxy Zoo 2 \citep[GZ2;][]{Willett2013} and NA10 \citep{Nair2010}. For the GZ2 classification, we have used two different methods to determine the bar frequency. Only about face-on galaxies ($i<30^{\circ}$) are selected. The dashed line and dotted line indicate the ridge of the star-forming main sequence and the passive sequence, respectively. The error bars are standard errors for the mean values or binomial errors for the fractions with 1$\sigma$ uncertainty. These values are calculated in the sliding box of 0.3\,dex in sSFR.}
 \label{fig:bar}
\end{figure}


\section{Results and Discussion}

As shown in Figure \ref{fig:gas}, we reprise the key result of \citetalias{Zhang2019} by replacing the x-axis with sSFR, instead of using SFR, to better minimize the $M_*$-dependence of our results. Figure \ref{fig:gas} shows that during the quenching of the massive central disk galaxies, as their sSFR decreases, their \hi~gas mass remains surprisingly constant, but their molecular gas mass and star formation efficiency drop rapidly. The depletion of molecular gas possibly begins from the central region of galaxies, since the suppression of star formation in massive galaxies is an inside-out process as shown by the results from color gradient and integral field unit (IFU) spectroscopy surveys \citep[e.g.][]{LiCheng2015,Pan2015self,Ellison2018self,Guo2019self,Lin2019self}. The dust content provides an alternative estimate of the gas mass in galaxies \citep[e.g.,][]{Leroy2011self,Yesuf2019self,Piotrowska2020}. The low value of dust attenuation within the SDSS fiber as shown in the right panel of Figure \ref{fig:gas} infers that there is little cold gas left in the central region of disk galaxies undergoing the quenching process.

In this section, we perform an analysis on the internal properties, such as structural parameters, bar frequency, and AGN frequency, for these galaxies to explore the internal quenching mechanisms. We follow the same sample selection criteria of \citetalias{Zhang2019}, i.e., only select massive central disk galaxies in the fixed stellar mass bin $10^{10.6}$$-$$10^{11}\Msol$. Here we use a narrow stellar mass bin to minimize the dependence of the various galaxy properties on stellar mass. Most of the plots in this paper will follow the same fashion as Figure \ref{fig:gas}, i.e., plotted as a function of continuously decreasing sSFR to indicate the quenching process.

\subsection{Structure}
\subsubsection{Compaction and Bulge Buildup}

To minimize the effect of dust attenuation and obtain more accurate visual classification of morphology, we only select nearly face-on galaxies with inclination angle $i<30^{\circ}$ in the study of the internal structure in this section. The disk inclination angles used here are from the fitting of \citet{Simard2011}. The top three panels of Figure \ref{fig:str} show the change of key structural parameters of face-on central disk galaxies as a function of sSFR. From main-sequence galaxies to the galaxies with lowest sSFR, the mean central velocity dispersion ($\sigma_*$) increases (top left panel), the average size of the galaxies measured by the half-light radii $R_{50}$ decreases (top middle panel), and the average concentration measured by $R_{90}$/$R_{50}$ increases (top right panel). The change of these three parameters indicates that central disk galaxies on average become smaller and more concentrated in the quenching process. For the galaxies above the main sequence (log sSFR/yr$^{-1}>-10.3$), galaxies also tend to be smaller and concentrated, which is consistent with previous works \citep[e.g.][]{Wuyts2011,Morselli2017self}. However, a detailed discussion about these high-sSFR galaxies is not the focus of this paper.

The change of $\sigma_*$, $R_{50}$, and $R_{90}$/$R_{50}$ shown by the top three panels of Figure \ref{fig:str} indicates that disk galaxies in the process of being quenched on average are more centrally concentrated compared to star-forming disk galaxies. Thus, we further decompose these central disk galaxies into their bulge component and disk component using the catalogs of \citet{Simard2011} and \citet{Mendel2014}. The results from the fitting model with a pure exponential disk and a de Vaucouleurs bulge (S\'ersic index $n_{b\rm }$ = 4) are used. The total stellar mass from the sum of $M_{\rm bulge}$ and $M_{\rm disk}$ agrees very well with the mass directly derived from the $k$-correction code, with a small scatter of $\sim$0.1 dex on average. We define the value of $M_{\rm bulge}/(M_{\rm bulge}+M_{\rm disk}$) as the bulge-to-total mass ratio ((B/T)$_{\rm mass}$) for our sample. About 11\% of disk galaxies that cannot be robustly decomposed into bulge+disk systems \citep[e.g. $\vert\Delta$(fiber color)$\vert>0.116$ or $P_{\rm ps}>0.32$; see details in][]{Simard2011} are not included in our analysis. Including these galaxies produces very small changes to the results.

The bottom left panel of Figure \ref{fig:str} shows that the average (B/T)$_{\rm mass}$ increases with decreasing sSFR, which clearly indicates that disk galaxies that undergoing quenching have a more massive bulge compared to star-forming ones. Meanwhile, the effective radius ($R_{\rm e}$) of the bulge component decreases with decreasing sSFR as shown in the bottom middle panel of Figure \ref{fig:str}. This hence requires the bulge to become more compact. Indeed, as shown in the bottom right panel of Figure \ref{fig:str}, the bulge's average surface mass density within its effective radius ($\mu_*$ = $M_*/(2\pi R_{\rm e}^2)$) increases accordingly during quenching. Interestingly, the effective radius and surface density of the disk component change weakly with sSFR, i.e. during the quenching process, the disk component remains largely unchanged. The fact that disk galaxies become more compact (top three panels of Figure \ref{fig:str}) is mainly driven by the growth and compaction of the bulge component as seen in the bottom three panels of Figure \ref{fig:str}.

\subsubsection{Bar Frequency}

Figure \ref{fig:bar} demonstrates the change of bar frequency in the quenching process for these central disk galaxies within the stellar mass range of $10^{10.6}-10^{11}\Msol$. We selected barred galaxies from two independent works, both based on visual classification of SDSS galaxies: Galaxy Zoo 2 \citep[GZ2;][]{Willett2013} and NA10 \citep{Nair2010}. We used two different methods to determine the bar frequency in GZ2. In GZ2, each galaxy has a large number of independent inspections by volunteers. For a given galaxy, if the debiased vote fraction of bar ($P_{\rm bar}$) is larger than 0.5, we define this galaxy as a barred galaxy. We apply this criterion to all our galaxies, and the derived bar frequency is plotted as the light-blue line in Figure \ref{fig:bar}. Alternatively, instead of determining whether an individual galaxy is a barred galaxy or not, we can directly calculate the average value of $P_{\rm bar}$. This is plotted as the dark-blue line in Figure \ref{fig:bar}. In the NA10 sample, galaxies have been inspected visually by experts, and it has been determined whether each galaxy contains a bar or not. The bar frequencies determined from these three different methods match very well. As the sSFR decreases, the bar frequency of disk galaxies increases dramatically from 30\% on the main sequence to larger than 80\% for the galaxies with lowest sSFR.

\begin{figure*}[t]
    \begin{center}
       \includegraphics[width=175mm]{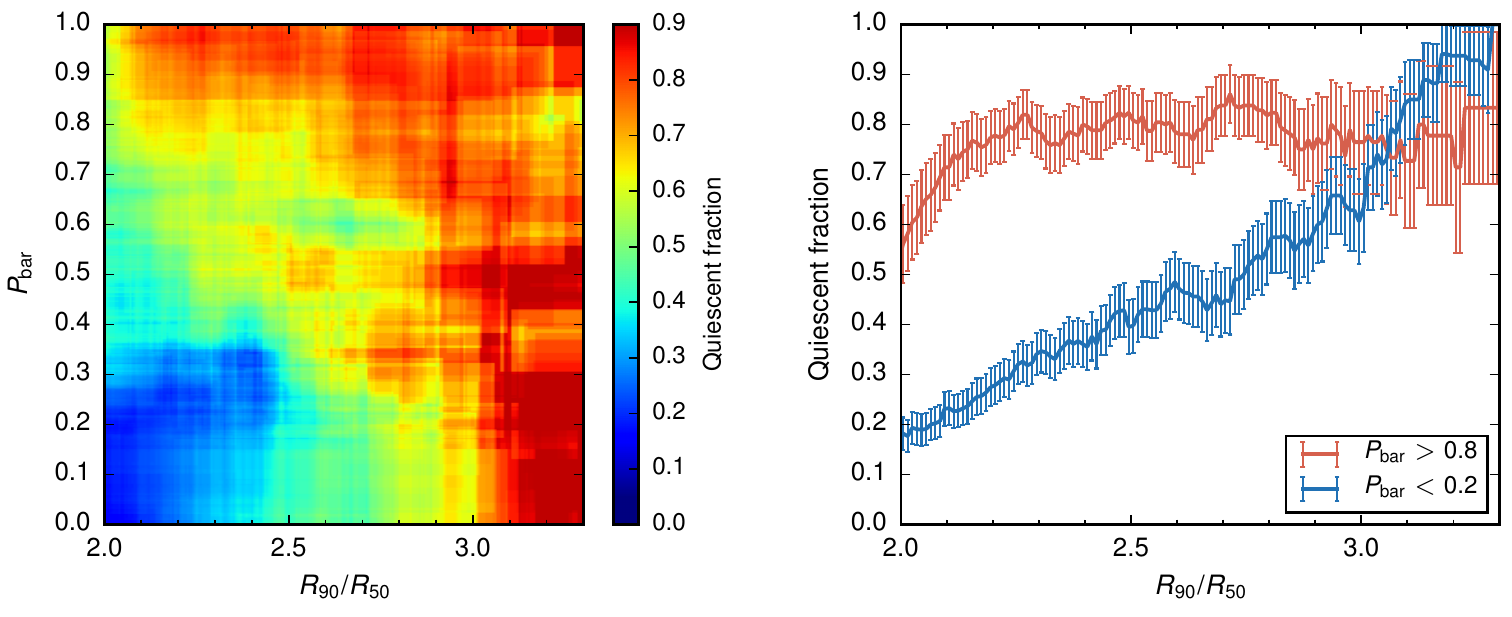}
    \end{center}
\caption{\textbf{Left:} The quiescent fraction as a function of $R_{90}$/$R_{50}$ and $P_{\rm bar}$ for massive central disk galaxies in the stellar mass range of $10^{10.6}$$-$$10^{11}\Msol$. The quiescent fractions are determined within moving boxes of size 0.4 in $R_{90}$/$R_{50}$ and 0.3 in $P_{\rm bar}$. As shown by the color-coding, the quiescent fraction increases to $>80$\% either for galaxies with $R_{90}$/$R_{50}>3.0$ or for galaxies with $P_{\rm bar}>0.8$. \textbf{Right:} The quiescent fraction for credible barred galaxies ($P_{\rm bar}>0.8$; in red line) and credible unbarred galaxies ($P_{\rm bar}<0.2$; in blue line) as a function of $R_{90}$/$R_{50}$ for this sample. The quiescent fractions are calculated in the sliding box of 0.4 in $R_{90}$/$R_{50}$ and the error bars are binomial errors with 1$\sigma$ uncertainty. In these two panels, galaxies with log sSFR/yr$^{-1}<-11$ are defined as quiescent ones. Only about face-on galaxies ($i<30^{\circ}$) are selected for this plot. }
 \label{fig:2d}
\end{figure*}

\subsubsection{The Role of Bulge and Bar in Quenching Star Formation}

The results shown in Figure \ref{fig:str} are consistent well with the conventional picture that massive galaxies become more concentrated in the process of star formation quenching \citep[e.g.][]{Cheung2012self,Wake2012self, Bluck2014self,Belfiore2017self}. However, we reiterate that the sample analyzed in our work only contains galaxies with clear stellar disks. The galaxies classified as ellipticals or uncertains in GZ are excluded in our analysis. Thus, the trends shown in Figure \ref{fig:str} are not caused by the increasing fraction of elliptical galaxies when sSFR decreases.

In simulation, the compaction and bulge buildup in disk galaxies can be triggered by mergers \citep{Barnes1991self,Hopkins2010self} and inflow driven by violent disk instabilities \citep{Dekel2014}. The massive central bulge can stabilize the surrounding gas disk against fragmentation to bound clumps \citep{Martig2009,Gensior2020}. \citet{Dekel2020a,Dekel2020b} predict that the gas ring around a post-compaction massive bulge can be stabilized against shrink when the mass ratio of the gas ring to the central bulge is well below unity. Thus, the significant growth of the bulge component in the quenching process shown in Figure \ref{fig:str} could be responsible for the low star formation efficiency in quiescent disk galaxies as shown in \citetalias{Zhang2019} and previous works \citep[e.g.,][]{Saintonge2012self,Oemler2017,Tacconi2018self,Piotrowska2020}. It should be noted that we do not suggest that the compaction and bulge buildup happen during the quenching process in the local universe. The progenitors of the local passive galaxies should be the star-forming galaxies at redshift $z\sim0.5$ or higher \citep{Peng2015}, so the compaction and bulge buildup may happen at high redshift. When the bulge became massive enough, it started to contribute to suppress star formation efficiency as we mentioned above. The quenching process may take several gigayears or even longer and then produce the trend of the B/T$-$sSFR relation that we observed at local universe (Figure \ref{fig:str}).

The galactic bar may also play an important role in quenching star formation in disk galaxies. By focusing on a clean and mass-completed sample that undergoing mass quenching, i.e., face-on central disk galaxies within a narrow stellar mass range, Figure 4 shows that the frequency of a strong bar increases significantly for central massive disk galaxies undergoing quenching, in agreement with previous works \citep{Masters2010self,Cheung2013,Fraser2018,Guo2020self}. However, we cannot determine the causal effect of the bar, in the sense that the bar can quench star formation \citep[e.g.,][]{George2020} or the bar instability is more likely to develop in quenched, molecular gas poor galaxies \citep[e.g.,][]{Shen2010}. In hydrodynamical simulations, it is clearly shown that the bar can drive gas within the co-rotation radius of the bar into the center of the galaxy and enhance the nuclear star formation, which accelerates the gas consumption in the galaxy and produces a gas-depleted region on the scale of several kiloparsecs \citep{Athanassoula2013,Spinoso2017,Khoperskov2018self}. This bar quenching scenario is supported by the observations of the gas distribution in some nearby barred galaxies \citep{Sheth2000self,Sheth2005,George2019M95,George2020,Newnham2020}. The significant increase of the bar frequency as sSFR decreases shown in Figure \ref{fig:bar} provides further observational statistical evidence that the galactic bar may contribute to star formation quenching.

In principle, the formation of a massive bulge and of a galactic bar can both contribute to star formation quenching. Indeed, on average, both of bulge mass and bar frequency increase when the sSFR decreases in massive disk galaxies in our sample. Hence, the question is whether these two components are intrinsically correlated or whether they give rise to two independent quenching channels. To investigate this question, we plot the fraction of quiescent galaxies (defined as galaxies with log sSFR/yr$^{-1}<-11$) as a function of $P_{\rm bar}$ and $R_{90}$/$R_{50}$ for our sample, as shown in the left panel of Figure \ref{fig:2d}. Here we use the values of concentration index $R_{90}$/$R_{50}$, instead of using B/T, to indicate the growth of bulge, since the values of B/T may be inaccurate without considering the bar component in the decomposition procedure \citep{Gao2017self,Morselli2017self}. As shown in Figure \ref{fig:str}, both of the quiescent and starburst galaxies have higher values of $R_{90}$/$R_{50}$. Since here we are focusing on the quenching process, we exclude the galaxies with elevated star formation above the main sequence (log sSFR/yr$^{-1}>-10.3$) in the calculation of quiescent fraction. As shown by the color-coding in the left panel of Figure \ref{fig:2d}, the fraction of quiescent galaxy increases to $>80$\% either for galaxies with $R_{90}$/$R_{50}>3.0$ or for galaxies with $P_{\rm bar}>0.8$.

One should note that the value of $P_{\rm bar}$ from GZ2 cannot be directly used as a proxy of any physical quantity, such as bar length or bar strength. A greater value of $P_{\rm bar}$ of a galaxy indicates that it is more likely to be a barred galaxy. Thus, we define a credible barred galaxy sample by $P_{\rm bar}>0.8$ and a credible unbarred galaxy sample by $P_{\rm bar}<0.2$. About 65\% of our sample belongs to these two categories. The right panel of Figure \ref{fig:2d} shows the quiescent fraction for these two credible samples as a function of $R_{90}$/$R_{50}$. For unbarred galaxies, the quiescent fraction gradually goes up with increasing $R_{90}$/$R_{50}$ as shown by the blue line. In contrast, the barred galaxies have a high and almost constant quiescent fraction of $\sim$80\%, largely independent of their concentration. The quiescent fraction drops slightly at the left end with the lowest concentration, but it is still larger than 50\%. As argued by \citet{Cheung2013}, the existence of quenched disk galaxies that host only pseudo-bulges would be a strong evidence that the bar can cause quenching without the effect of classical bulges. Figure \ref{fig:2d} shows that most barred galaxies have been quenched (or undergoing quenching) even for galaxies with $R_{90}$/$R_{50}<2.3$, which corresponds to the typical value for galaxies with only pseudo-bulges according to the statistic of \citet{Gadotti2009self}. The results of Figure \ref{fig:2d} suggest that the quenching process due to bulge and bar can be separated in massive disk galaxies.

Although the effect of quenching due to bulge and bar is different, the growth of the bulge and bar could be closely correlated, since the buildup of the central mass can be efficiently triggered by the bar. A strong bar can drive gas to the center of galaxies \citep{Athanassoula1992,Sheth2005,Spinoso2017,George2020} and therefore enhance the nuclear star formation \citep{Coelho2011,Lin2016,Robichaud2017} to build up a pseudo-bulge \citep[see][for a review]{Kormendy2004self}. Therefore, the quenching effect of bulge and bar may in fact work in tandem in disk galaxies. While the global star formation is gradually suppressed with the formation of a massive bulge, the formation of the galactic bar accelerates the secular evolution of the galaxy by driving gas inflow into the central region and helps to build the central mass. Then, the massive bulge can further stabilize the gas in the outer disk and reduce the star formation efficiency. The resolved observations of cold gas distribution and precise structure decomposition and kinematics measurement for a large sample will be essential for further understanding the role of bulges and bars in quenching star formation.

\begin{figure*}[ht]
    \begin{center}
\includegraphics[width=180mm]{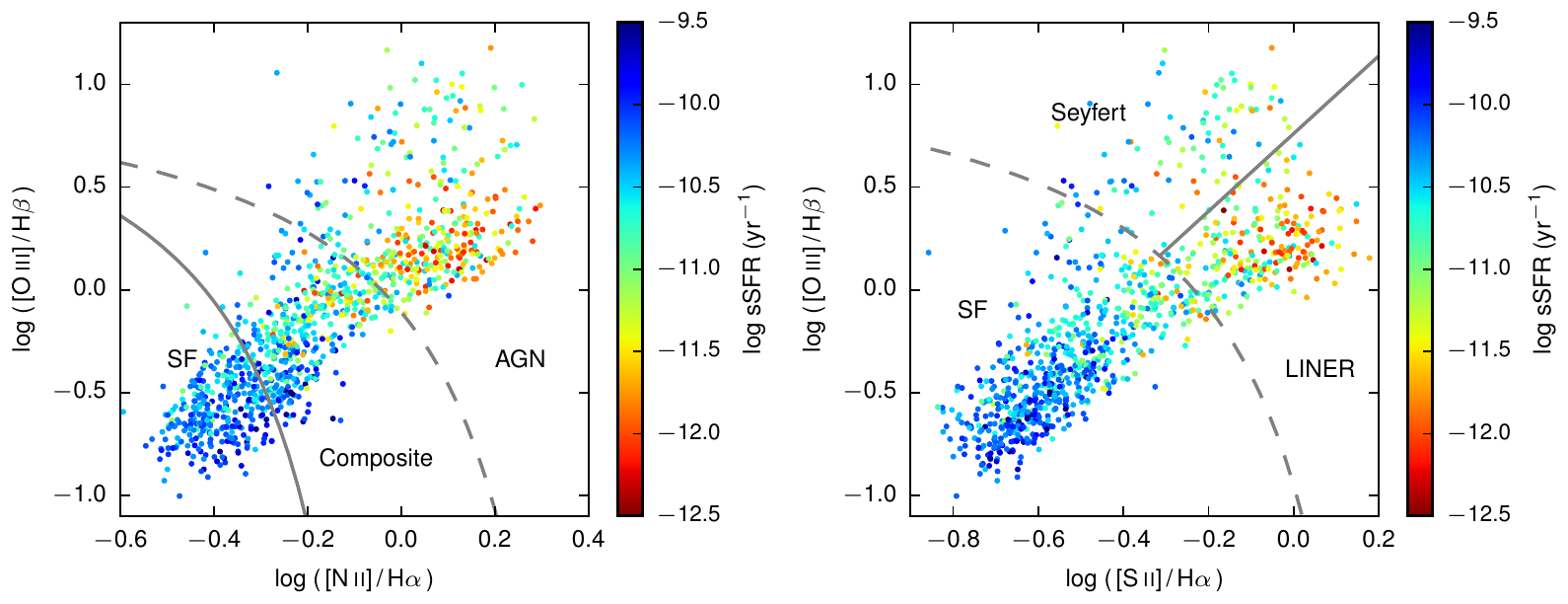}
   \end{center}
\caption{\textbf{Left:} BPT diagram for central disk galaxies within the stellar mass range of $10^{10.6} - 10^{11}\Msol$. The solid line shows the empirical division between AGNs and star-forming (SF) galaxies \citep{Kauffmann2003c}. The dashed line indicates the theoretical upper limit for pure starburst models, and galaxies above this line are classified as AGNs \citep{Kewley2001}. \textbf{Right:} Seyfert/LINER classification. The solid line is the empirical division between LINERs and Seyferts \citep{Kewley2006}. In both panels, the central disk galaxies with S/N $>3$ in all emission lines in the stellar mass range of $\sim 10^{10.6} - 10^{11}\Msol$ are plotted and color-coded according to their sSFR.}
 \label{fig:bpt}
\end{figure*}

\begin{figure*}[ht]
    \begin{center}
\includegraphics[width=180mm]{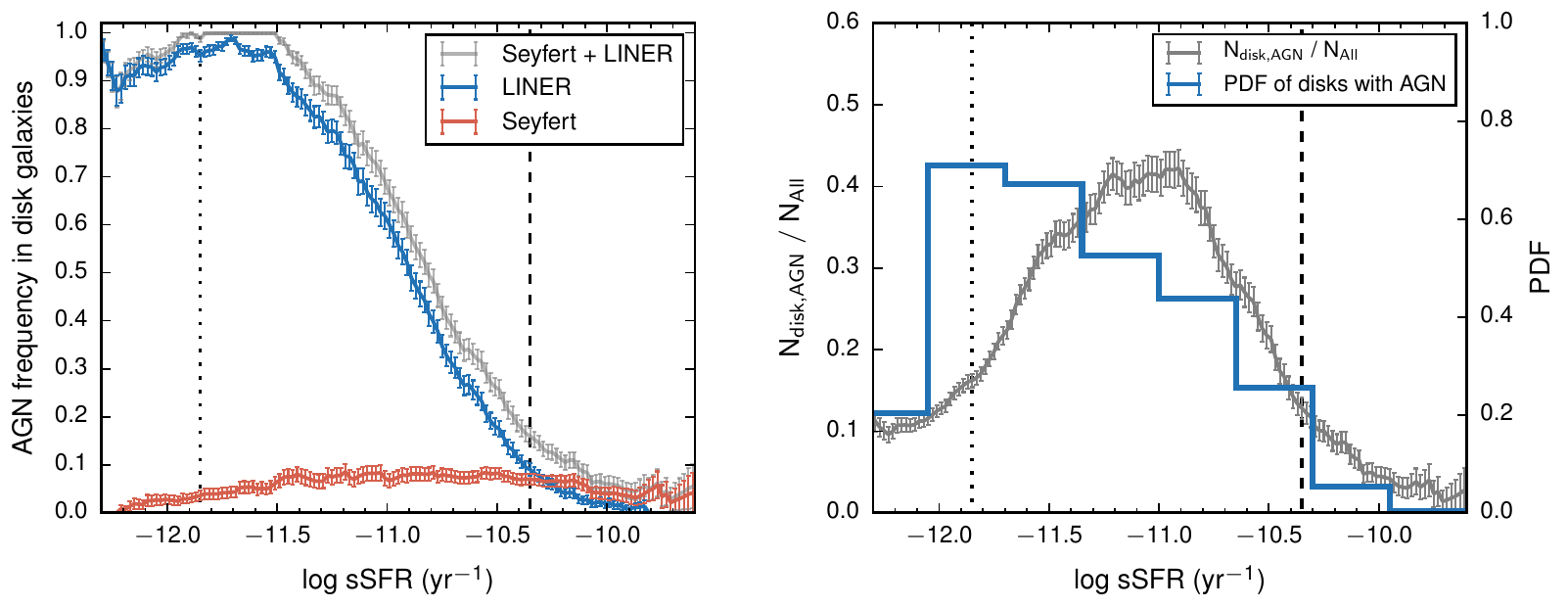}
   \end{center}
\caption{\textbf{Left:} The frequency of all AGNs (gray line), LINERs (blue line), and Seyferts (red line) in the central disk galaxies as a function of sSFR, within the stellar mass range of $10^{10.6}$$-$$10^{11}\Msol$. \textbf{Right:} The ratio of the number of central disk galaxies with AGNs (including both LINERs and Seyferts) to the number of all central galaxies (including all morphology types) as a function of sSFR (gray line) in the same stellar mass range. This fraction is apparently peaked right in the green valley. The blue line shows the probability density function (PDF) of central disk galaxies with AGNs, which continuously increases with decreasing sSFR and reaches its maximum value at the passive sequence. In each panel, the fractions are calculated in the sliding box of 0.5\,dex in sSFR, and the error bars are binomial errors with 1$\sigma$ uncertainty. The dashed line and dotted line indicate the ridge of the star-forming main sequence and the passive sequence, respectively.}
 \label{fig:agn}
\end{figure*}

\subsection{Active Galactic Nucleus}  \label{sec:agn}

\subsubsection{AGN Frequency}   \label{sec:fagn}

AGN feedback has often been proposed as the main mechanism responsible for quenching massive galaxies \citep{Croton2006,Fabian2012,Harrison2017a}. To investigate the interrelationship between AGN feedback and star formation quenching, we identify AGNs by the BPT diagnostic diagram \citep{Baldwin1981}. As shown in the left panel of Figure \ref{fig:bpt}, galaxies above the upper limit of pure starburst models \citep[dashed line;][]{Kewley2001} are selected as AGNs. We further classify AGNs into Seyferts and low-ionization nuclear emission-line regions (LINERs) by the empirical division defined by \citet{Kewley2006} as shown by the solid line in the right panel of Figure \ref{fig:bpt}. As in \citet{Brinchmann2004}, about 20\% of our sample is classified as low-S/N AGNs (galaxies that have \nii $\lambda$6584/\ha~$>0.6$ and S/N $>3$ in both lines, but their \oiii $\lambda$5007~and/or \hb~have too low S/N to be useful). These low S/N AGNs are classified into the LINER class, since the values (or upper limits) of their Eddington ratios (see details in Section \ref{sec:Edding}) are similar to those of LINERs but are much smaller compared to those of Seyferts.

The frequency of LINERs decreases significantly with redshift (as shown in Figure \ref{fig:agnz} in the Appendix). This is an apparent observational selection effect due to the fact that the emission lines used to classify LINERs become progressively too faint to be detected and more stellar emission is involved in the large aperture of the SDSS fiber (3\arcsec) at higher redshifts. We describe the correction of this bias in the Appendix, and we only use the galaxies in the redshift range of 0.02$-$0.05 in this section to diminish this effect. The aperture of the SDSS fiber corresponds to 0.6$-$1.5\,kpc in radius in this redshift range.

In the left panel of Figure \ref{fig:agn}, we show the frequency of AGNs (including both Seyferts and LINERs) in central disk galaxies as a function of sSFR. From star-forming galaxies to the ones with the lowest sSFR, the AGN frequency increases dramatically from 10\% to almost 100\% as shown by the gray line. This figure also shows that the AGNs in star-forming disk galaxies are dominated by Seyferts (red line), and those in the disk galaxies with the lowest sSFR are dominated by LINERs (blue line). Therefore, the trend of increasing AGN fraction with decreasing sSFR is caused by the increasing fraction of LINERs. It has been known that the LINER emission detected in large apertures is not a reliable indicator of AGN activity \citep{Ho2008}, and it could be excited by the hard radiation field produced by evolved hot stars \citep[e.g.,][]{Stasinska2008,Yan2012,Belfiore2016c}, hence is rephrased as ``LIERs'' by \citet{Belfiore2016c}. In this case, the nearly 100\% LINER frequency in quiescent disk galaxies may be a consequence of star formation quenching since the \hii~regions in these galaxies are going to be extinct. However, the trends of increasing B/T (and hence bulge mass) and increasing central velocity dispersion as shown in Figure \ref{fig:str} indicate more massive central black holes in quenched disk galaxies or disk galaxies that are in the process of being quenched. As discussed by \citet{Ho2008}, at least the nuclear emission in many of these LINER-like galaxies is truly associated with low-luminosity AGNs. On the other hand, most LINERs were found to contain point X-ray cores, which indicates that their central black holes are indeed active \citep{Ho2001self, She2017self}.

Another important aspect of AGNs is that, as shown in the right panel of Figure \ref{fig:agn}, the ratio of the number of central disk galaxies with AGNs to the number of all central galaxies (including all morphology types) as a function of sSFR (gray line) is apparently peaked right in the green valley. This concentration of disk AGN host galaxies in the green valley has often been quoted as the key evidence that AGNs quench galaxies \citep{Nandra2007,Schawinski2007,Leslie2016,Silverman2019}. The blue line shows the probability density function (PDF) of the disk galaxies with AGNs as a function of sSFR, which continuously increases with decreasing sSFR and reaches its maximum value for galaxies with the lowest sSFR. This is the place where essentially every disk galaxy hosts an AGN, as shown in the left panel. This suggests that if AGN feedback is contributing to quenching disk galaxies, it should be a continuous process, not only operating in the green valley. The drop of the AGN fraction (gray line) in the low-sSFR regions is due to the low AGN frequency in elliptical galaxies.

\subsubsection{Low-luminosity AGN Feedback} \label{sec:Edding}

To estimate the Eddington ratios of these Seyferts and LINERs in central disk galaxies, we use $\sigma_*$ to estimate the black hole masses from the $M_{\rm BH}$-$\sigma_*$ relation given by log($M_{\rm BH}$/\Msol)  = 5.20 log($\sigma_*$/200 km s$^{-1}$) + 8.32 \citep{She2017self}. The Eddington luminosities are calculated by $L_{\rm Edd}$ = (1.3 $\times10^{38}~{\rm erg~s^{-1}}) M_{\rm BH}/\Msol$. Then, we use \oiii$\lambda$5007 to estimate the bolometric luminosities by using $L_{\rm bol}/(\rm10^{40}~erg~s^{-1}) = 112\left(L_{\rm [O\,III]}/(\rm 10^{40}~erg~s^{-1})\right)^{1.2}$, which is a power-law fit to the luminosity-dependent bolometric corrections found by \citet{Lamastra2009} and parameterized by \citet{Trump2015self}. The general trend of our results will not change using other bolometric corrections \citep[e.g.,][]{Heckman2004self,Kauffmann2009self}. The \oiii$\lambda$5007 luminosities have been corrected for internal extinction from the observed Balmer decrement, assuming an intrinsic value of \ha/\hb~= 3.1 and the extinction curve of \citet{Cardelli1989self}.

\begin{figure}[t]
    \begin{center}
       \includegraphics[width=\columnwidth]{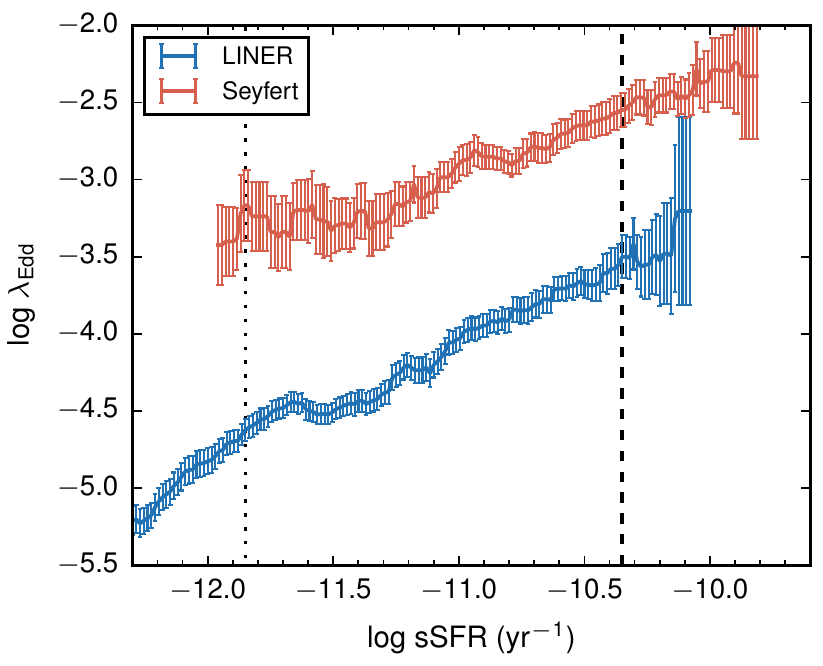}
    \end{center}
\caption{Average Eddington ratio ($\lambda_{\rm Edd}$) of LINERs (blue line) and Seyferts (red line) in the central disk galaxies as a function of sSFR, within the stellar mass range of $10^{10.6}$$-$$10^{11}\Msol$. The average values are calculated in the sliding box of 0.5\,dex in sSFR, and the error bars are standard errors with 1$\sigma$ uncertainty. The dashed line and dotted line indicate the ridge of the star-forming main sequence and the passive sequence, respectively.}
 \label{fig:agn_edd}
\end{figure}

The average Eddington ratio ($\lambda_{\rm Edd}=L_{\rm bol}/L_{\rm Edd}$) of LINERs and Seyferts in the central disk galaxies as a function of sSFR is shown in Figure \ref{fig:agn_edd}. For LINERs that are dominant in quenched or quenching disk galaxies, the mean Eddington ratio is around $10^{-5}-10^{-4}$, which is well consistent with the statistics from the Palomar survey \citep{Ho2009self}. It should be noted that some of the LINER emission in quenched galaxies could be powered by non-AGN sources (e.g., hot evolved stars). Thus, the Eddington ratios of LINERs in our work should be regarded as upper limits of the true values.

\begin{figure*}[t]
    \begin{center}
       \includegraphics[width=180mm]{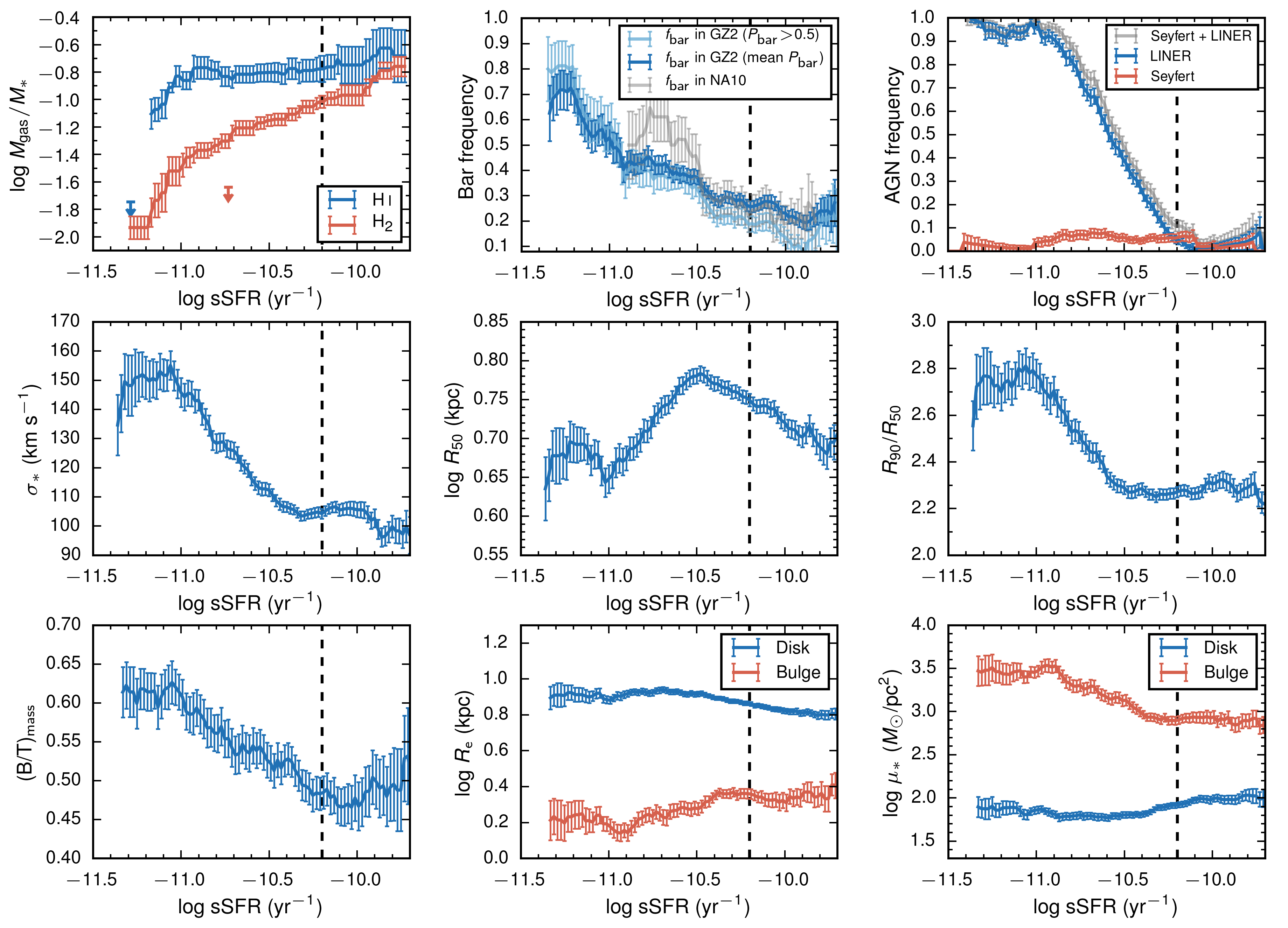}
    \end{center}
\caption{Gas-to-stellar mass ratio, bar frequency, AGN frequency, and key structural parameters as a function of sSFR for central disk galaxies within the stellar mass range of $10^{10.6}$$-$$10^{11}\Msol$. The SFRs used in this plot are derived from the SED fitting of UV, optical, and mid-IR bands \citep[GSWLC-M2 catalog,][]{Salim2016a,Salim2018self}. The dashed line indicates the ridge of the star-forming main sequence defined in \citet{Salim2018self} by using those SED-based SFRs. All labels are the same as those in the previous corresponding plot in this paper.
}
 \label{fig:SFR}
\end{figure*}

If AGN feedback is indeed the dominant quench mechanism over ``non-AGN'' solution mechanisms as suggested by latest hydrodynamic simulations \citep{Su2019self}, it should operate by preventing further gas cooling and gas inflow \citep[``preventive feedback''; see the review of ][]{Husemann2018}. The remaining molecular gas can be consumed by star formation or expelled by the wind from the AGN, or by both. Observations also show that the radio-loudness increases with decreasing Eddington ratio \citep{Ho2002self,Ho2008,Ho2009self}. The very low Eddington ratio of the LINERs as shown in Figure \ref{fig:agn_edd} hence suggests that the primary AGN feedback mode is likely kinetic-mode feedback, when winds and jets become dominant. Indeed, the latest numerical simulations find that uncollimated winds produced by low-luminosity AGNs can inject sufficient amount of energy and momentum into the surrounding medium and push the cold gas to the outer region of the galaxy \citep{Yuan2014,Yuan2015a,Bu2016,Sadowski2016,Weinberger2018self,Yuan2018self,Terrazas2020}.

Recent observations show that the local AGN host galaxies have similar \hi~gas content compared to non-AGN hosts when other parameters are matched \citep{Ellison2019}. This is consistent with our findings that the disk galaxies in the process of being quenched have the same large \hi~gas reservoir as star-forming galaxies but sharply increasing AGN frequency (Figure \ref{fig:gas} and \ref{fig:agn}). The existence of a regularly rotating \hi~disk around the galaxies that are undergoing quenching also puts strong constrains on the strength and geometry of AGN feedback in the sense that the feedback cannot be too violent on the disk plane; otherwise, the \hi~disk could not survive. In the latest IllustrisTNG simulation \citep{Weinberger2017,Weinberger2018self}, the quasar-mode feedback with high black hole accretion rate cannot quench galaxies, but the quenching of massive central galaxies happens coincidentally with kinetic-mode feedback. And in IllustrisTNG, the SFR of galaxies will significantly decrease when the accumulated black hole wind energy exceeds the gravitational binding energy of the gas within galaxies and it occurs at a particular black hole mass threshold above which the feedback switches from thermal to kinetic injection \citep{Davies2019self,Terrazas2020}. The low Eddington ratio and large black hole mass of the quiescent galaxies and galaxies that are undergoing quenching in the IllustrisTNG simulation are consistent well with the observational results we find here.

\subsection{ The Effect of using Alternative SFR Estimator} \label{sec:SFR_Salim}

As discussed in \citetalias{Zhang2019}, using different SFR estimators may produce different results. \citet{Cortese2020} show that utilizing the SFRs calculated from UV and IR photometry will lead to different conclusions. To address this concern, we repeat all analysis in this paper by using the SFRs derived from the SED fitting of UV, optical, and mid-IR bands \citep[GSWLC-M2 catalog;][]{Salim2016a,Salim2018self}. Only 49\% of our sample is included in the GSWLC-M2 catalog, which is due to the limited sky coverage of the GALEX medium-deep UV survey. For this 49\% of our sample, their SED-based SFRs are calculated regardless of whether they have detections in UV or IR, or both. Hence, this should not introduce a bias compared to the MPA-JHU sample SFRs, except that the sample size becomes smaller. We show in Figure \ref{fig:SFR} the gas-to-stellar mass ratio, bar frequency, AGN frequency, and key structural parameters for central disk galaxies within the stellar mass range of $10^{10.6}-10^{11}\Msol$ as a function of sSFR derived from those SED-based SFRs. To keep consistent with previous figures, the stellar masses used in Figure \ref{fig:SFR} are still obtained from the fits to the optical photometry by using the k-correction program as described in Section \ref{sec:SDSS}. The dashed line in each panel indicates the ridge of the star-forming main sequence defined in \citet{Salim2018self} using these SED-based SFRs.

As shown in Figure \ref{fig:SFR}, the lowest sSFR of the central disk galaxies reaches $\sim$1\,dex below the main sequence when using the SED-based SFRs, while it reaches $\sim$1.5\,dex below the main sequence using the MPA-JHU SFRs as shown in previous figures. Therefore, when the SED-based SFRs are used, there are few fully quenched, quiescent central disk galaxies in the local universe, regardless of their gas content. However, it should be noted that almost all of these disk galaxies with the lowest sSFR are LINERs as shown by Figure \ref{fig:agn} and the top right panel of Figure \ref{fig:SFR}. The UV and IR emission of these LINER galaxies can be contaminated by AGNs or old stellar populations, as currently considered for LINERs. Thus, their SFRs derived from the UV and IR photometry could be overestimated. We also note that, using this alternative SFR estimate, the upturn toward high sSFR has disappeared for the panels involving $\sigma_*$ and $R_{90}$/$R_{50}$, and to a certain extent also (B/T)$_{\rm mass}$. The detailed discussion of these starburst galaxies is not the focus of this paper.


Even if these galaxies can be defined as fully quenched or just quenching depending on the adopted SFR indicator, still the general trends below the star-forming main sequence of the plots shown in Figure \ref{fig:SFR} are fully consistent with those obtained by using the MPA-JHU SFR shown in previous figures. In fact, not only the general trends, but also at the lowest log sSFR, the absolute values of log $(M_{\rm H2}/M
_*)$ ($\sim$$-1.9$), bar frequency ($\sim$70\%), AGN frequency ($\sim$100\%), $\sigma_*$ ($\sim$150\,km/s), B/T ($\sim$0.6), and $R_{90}/R_{50}$ ($\sim$2.7) are the same for both SFR indicators. Therefore, the absolute amount of changes in these parameters during quenching are the same, independent of the SFR indicator.

Putting together, the key facts found in our work hold for both the MPA-JHU SFR and SED-based SFR as follows. For these massive central disk galaxies with the lowest sSFR, (1) they have the same large \hi~gas reservoir as the star-forming ones; (2) they have $\sim$10 times less \htwo~gas and significantly suppressed star formation efficiency than the star-forming ones; and (3) they have significantly higher B/T, higher bar frequency, and higher LINER frequency than the star forming ones. These features (2 and 3) make them distinct from the properties of the star-forming ones near the main sequence. Moreover, it remains to be seen to which extent the UV/mid-IR emission of these LINERs is due to recent star formation or to AGN or old stellar populations. Thus, although the precise category (``fully quenched'' or ``green valley'') of the disk galaxies with the lowest SFR depends on the SFR indicator, it is certain that they are below the star-forming main sequence and are undergoing the quenching process. We stress that the main focus of our work is to study the change of the properties of disk galaxies as a continuous function of decreasing sSFR, and explore the physical mechanisms that drive these changes. As discussed above, these results (both trends and absolute amount of changes) are independent of the SFR estimator. The exact name of the galaxies with the lowest sSFR, i.e., fully quenched or green valley (i.e. galaxies in the process of being quenched), is therefore less critical.

By analyzing the cold gas content of the MaGMA sample, \citet{Hunt2020} also found that the \hi~mass of massive galaxies does not vary with SFR, but their \htwo~mass decreases with decreasing SFR, which is in excellent agreement with the trends shown by \citetalias{Zhang2019}.
The $\sim$10 times less \htwo~gas and significantly suppressed star formation efficiency in the disk galaxies with the lowest sSFR indicate that they are indeed in the process of being quenched. This is also supported by the Fundamental Formation Relation (FFR) proposed in \citet{Dou2021}, which reveals that the molecular gas fraction and star formation efficiency of galaxies decrease with decreasing sSFR. On the other hand, the structures (bulge, bar) of these disk galaxies in the quenching process are very different from those of the star-forming galaxies. Thus, the majority of these low-sSFR disk galaxies cannot rejuvenate and return to the main sequence in their subsequent evolution.

\begin{figure*}[t!]
    \begin{center}
\includegraphics[width=170mm]{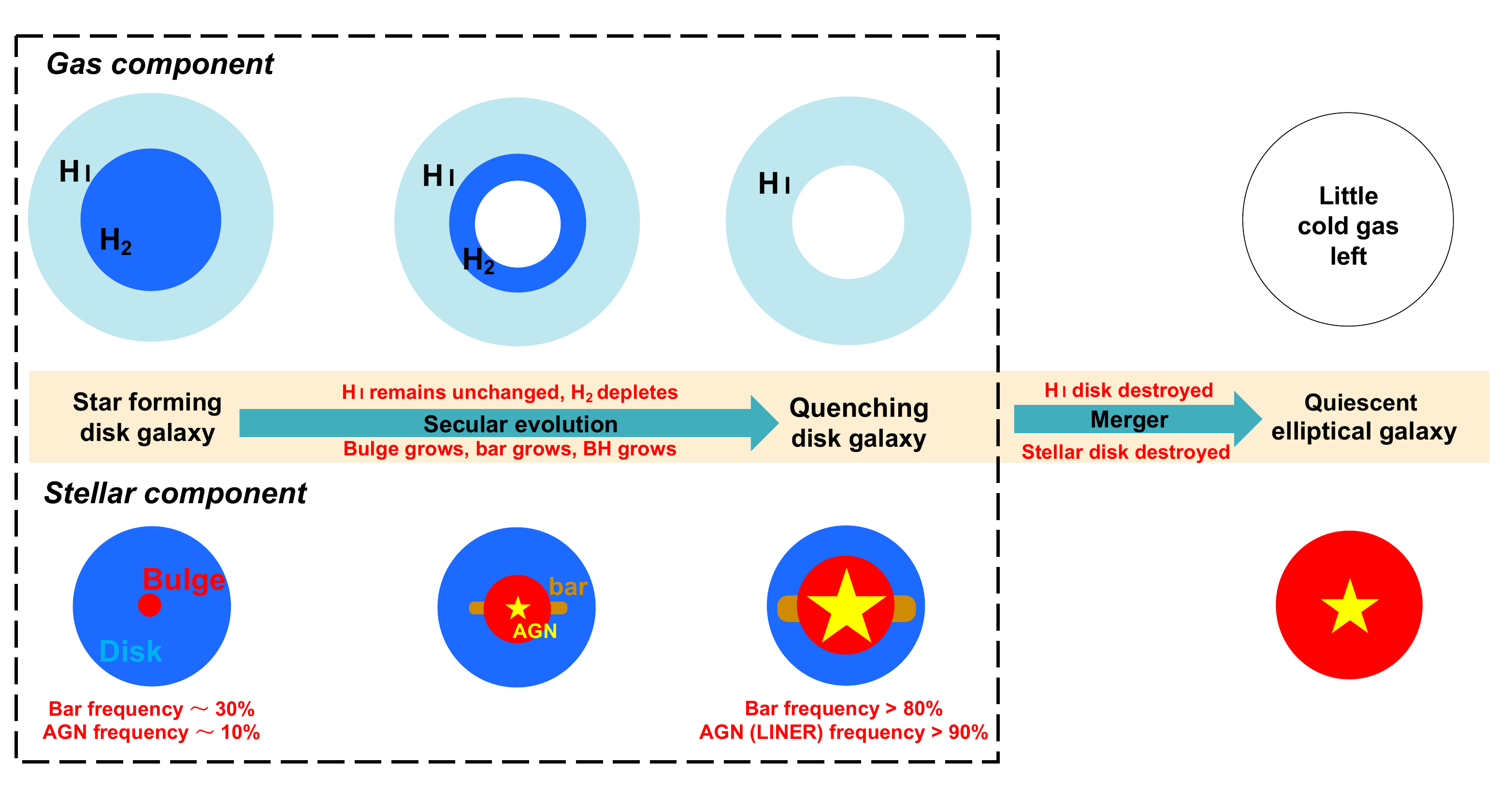}
    \end{center}
\caption{Schematic diagram of the detailed quenching process for massive central galaxies. The \hi~gas reservoir remains unchanged in the quenching process of disk galaxies, but the molecular gas is gradually depleted. On the other hand, disk galaxies become more concentrated owing to the significant growth of the bulge component. The stellar disk remains largely unchanged. The bar frequency in disk galaxies dramatically increases from 30\% to larger than 80\% in the quenching process. The frequency of AGNs (including LINERs) selected from the BPT diagram increases from 10\% to almost 100\%. Nearly all massive central disk galaxies with the lowest observed sSFR are LINERs. The quenched disk galaxies or disk galaxies that are undergoing quenching process are subsequently transformed into quiescent elliptical galaxies by mergers in a random fashion, which also destroys their \hi~disk and leaves the ellipticals with little cold gas.
}
\label{fig:quench}
\end{figure*}

\section{Summary}

To investigate the possible physical mechanisms of mass quenching, we focus on massive central disk galaxies within the stellar mass range of $10^{10.6}$$-$$10^{11}\Msol$ in SDSS, which compose a clean sample that undergoing the quenching process with an internal physical origin. We showed that in \citetalias{Zhang2019}, as the SFR decreases, the \hi~gas mass of massive central disk galaxies remains surprisingly constant, but their molecular gas mass and star formation efficiency drop sharply. In this work, we further studied the change of internal properties of these galaxies during the quenching process. The main results and the detailed mass quenching process revealed by these two works are illustrated as the schematic diagram in Figure \ref{fig:quench} and summarized as follows:

(i) As shown in Figure \ref{fig:gas} and \citetalias{Zhang2019}, during the quenching of the massive central disk galaxies, as their sSFR decreases, their \hi~gas mass remains surprisingly constant, but their molecular gas, dust, and star formation efficiency drops rapidly. The identical symmetric double-horn \hi~profiles indicate similar regularly rotating \hi~disks in both main-sequence disk galaxies and quenched disk galaxies, as shown in \citetalias{Zhang2019}. The depletion of molecular gas possibly begins from the central region of galaxies, since the suppression of star formation in massive galaxies is an inside-out process as shown by the results from recent IFU surveys. It is also consistent with the low value of dust attenuation within the SDSS fiber as shown in the right panel of Figure \ref{fig:gas}, which infers that there is little cold gas left in the central region of these quenched disk galaxies or disk galaxies that are in the process of being quenched. However, future interferometry observations are needed to confirm the cold gas distribution in these galaxies.

(ii) In the quenching process, massive central disk galaxies become more concentrated on average as shown by the increasing central velocity dispersion ($\sigma_*$), decreasing half-light radii ($R_{50}$), and increasing concentration index ($R_{90}$/$R_{50}$). By decomposing central disk galaxies into their bulge component and disk component, we show that their average bulge-to-total mass ratio (B/T) increases with decreasing sSFR. Meanwhile, their bulges become more compact in the quenching process as indicated by the smaller effective radius and larger surface mass density of the bulge. The effective radius and surface mass density of the disk component remain largely unchanged. The fact that galaxies become more compact is mainly driven by the growth and compaction of their bulge component (see Figure \ref{fig:str}). These observational results are consistent with the bulge quenching mechanisms proposed in theoretical works, such as the morphological quenching and gravitational quenching.

(iii) The bar frequency in massive central disk galaxies dramatically increases from 30\% to larger than 80\% in the quenching process (see Figure \ref{fig:bar}). With given $R_{90}$/$R_{50}$, the barred galaxies have a much higher quiescent fraction than unbarred galaxies. The quiescent fraction of the credible barred sample is $\sim$80\% even for galaxies with very low concentration (see Figure \ref{fig:2d}). This indicates that the bar may play an important role in quenching massive disk galaxies, as suggested by recent hydrodynamical simulations.

(iv) The frequency of AGNs selected from the BPT diagram increases from 10\% to almost 100\% in the quenching process (see Figures \ref{fig:bpt} and \ref{fig:agn}). Nearly all massive central disk galaxies with the lowest observed sSFR are LINERs with very low Eddington ratios of $10^{-5}-10^{-4}$ (see Figure \ref{fig:agn_edd}). As discussed in Section 3.2, if AGNs are contributing to quenching, we suspect that the primary AGN feedback is operating in preventive mode, through kinetic feedback from low-luminosity AGNs. Hydrodynamic simulations show that the kinetic energy from low-luminosity AGN feedback can sufficiently quench the star formation in massive galaxies.

\vspace{0.3cm}

All of the above processes, bulge growth, bar growth, and AGN feedback, can work together to quench the star formation in the central massive disk galaxies. We stress that these three processes could be intrinsically close correlated. The galactic bar can accelerate the secular evolution of the galaxy by driving gas inflow into the central region and help to build the bulge and feed the black hole. The massive bulge can stabilize the gas and prevent the gas to collapse to form stars and hence reduce the star formation efficiency. Meanwhile, the more massive bulge hosts more massive black holes in the center, and a more massive black hole means stronger kinetic feedback, as suggested by recent simulations \citep[e.g.,][]{Weinberger2017,Weinberger2018self,Yuan2018self,Davies2019self,Terrazas2020}. The kinetic feedback can expel the molecular gas out of the disk, which explains the lower molecular gas content in the central region of the galaxy. The feedback from the AGN can simultaneously inject energy into the circumgalactic medium and prevent further gas cooling into the galaxy \citep{Zinger2020self}, hence strangulating the gas supply of the galaxy \citep{Peng2015}.

The existence of the regularly rotating \hi~disk around the quenched disk galaxies or disk galaxies that are undergoing the quenching process (as shown in \citetalias{Zhang2019}) is likely due to two reasons. First, the inflowing gas with excess angular momentum can settle on a stable outer ring of neutral hydrogen for a long timescale \citep{Peng2020,Renzini2020} in the case that lacks any perturbation.
The gas surface density at the large radii can be very low and below the critical density for phase transition from \hi~to \htwo~\citep{Bigiel2008} and hence for star formation \citep[e.g.,][]{Lemonias2014,Lu2015,Tacconi2020review}. Second, the cold gas could be kept in the outer region of the galaxy owing to the ejective mode AGN feedback. The existence of the \hi~disk around quenched disk galaxies or disk galaxies in the process of being quenched also puts strong constrains on the strength and geometry of AGN feedback in the sense that the feedback cannot be too violent on the disk plane; otherwise, the \hi~disk could not survive.

Since the mass-quenching process, such as AGN feedback or excessive angular momentum discussed above, is mainly operating on the gas, not on the stars, it is not expected to change the kinematics of the stars. Indeed, as shown in \citet{WangBT2020}, the quenched disk galaxies have similar stellar kinematics to their star-forming counterparts. On the contrary, elliptical galaxies occupy distinct regions on the ellipticity$-$spin parameter plane compared to the disk galaxies, indicating that an additional mechanism beyond quenching is required to change the stellar kinematics and morphology, such as merging. During the internal secular evolution, these massive disk galaxies may transform into quiescent elliptical galaxies by major mergers in a random fashion, which may destroy their \hi~disk and leave the ellipticals with little cold gas. In future work, we will further investigate which mechanism, in particular AGN feedback, kinematics of the inflowing gas, formation of a massive bulge, and bar-induced activity, is the dominant mass quenching mechanism, as well as their causal relationships.

\acknowledgments

We thank Robert Kennicutt, Sandra Faber, Nick Scoville, Barbara Catinella, Renbin Yan, Jing Wang, Hua Gao, and Min Du for useful discussions. Y.P. acknowledges NSFC grant Nos. 11773001, 11721303, and 11991052, and the National Key R\&D Program of China grant 2016YFA0400702. L.C.H. acknowledges National Key R\&D Program of China grant 2016YFA0400702 and NSFC Grant Nos. 11473002 and 11721303. R.M. acknowledges ERC Advanced grant 695671 ``QUENCH'' and support by the Science and Technology Facilities Council (STFC). A.D. acknowledges support from the grants NSF AST-1405962, GIF I-1341-303.7/2016, and DIP STE1869/2-1 GE625/17-1. Q.G. acknowledges NSFC grant Nos. 11573033 and 11622325 and the Newton Advanced Fellowships. F.M. acknowledges support from the INAF PRIN-SKA 2017 program 1.05.01.88.04. D.L. acknowledges the NSFC grant Nos. 11690024 and 11725313. A.R. acknowledges support from an INAF/PRIN-SKA 2017 (ESKAPE-HI) grant. 

\clearpage
\appendix

\section{Statistical Correction of the AGN Frequency}
\begin{figure*}[h]
    \begin{center}
       \includegraphics*[width=180mm]{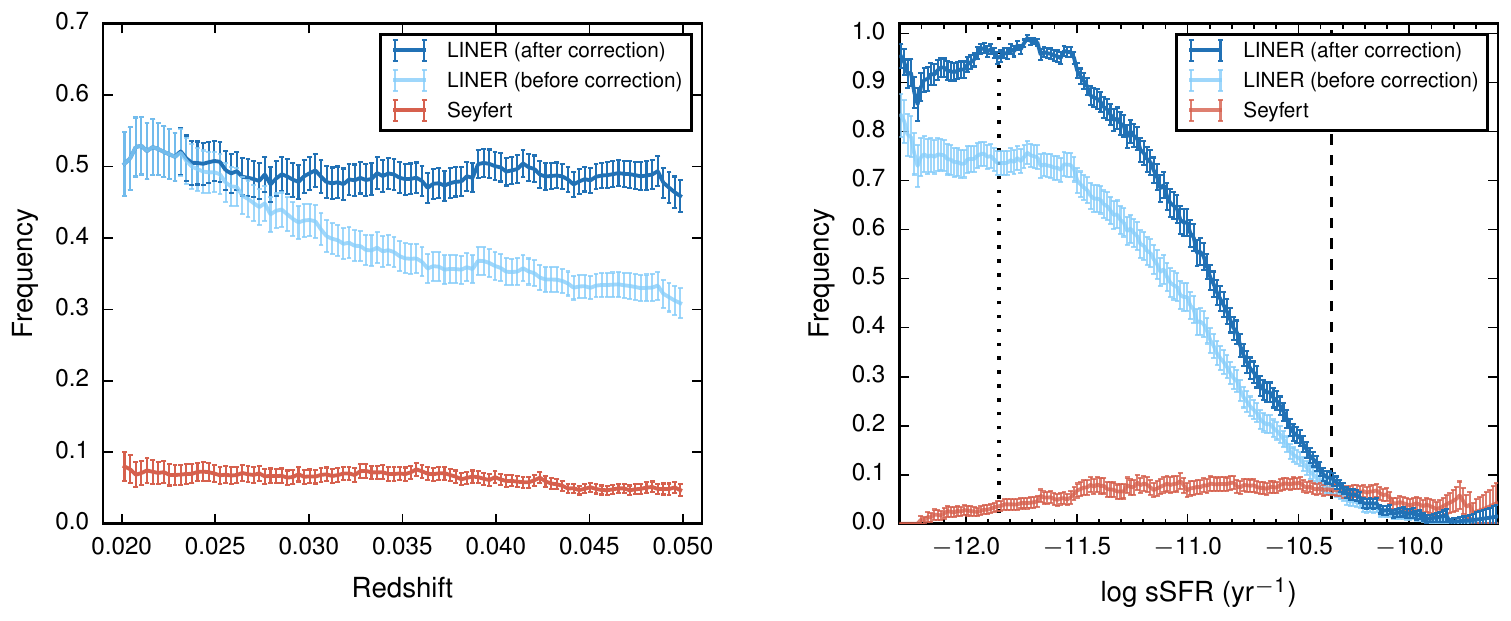}
    \end{center}
\caption {{\bf Left:} The frequency of LINERs (blue lines) and Seyferts (red line) for central disk galaxies in the stellar mass range of $10^{10.6}$$-$$10^{11}$\Msol
~as a function of redshift. The frequency of LINERs (light blue line) decreases significantly with redshift owing to the observational selection effect as explained in the text. The dark-blue line shows the LINER frequency after the statistical correction. {\bf Right:} The LINER frequency, before (light-blue line) and after (dark-blue line) the statistical correction, and Seyfert frequency (red line) for central disk galaxies as a function of sSFR in the same mass range. Error bars are derived from the binomial error of the fraction with 1$\sigma$ uncertainty in the sliding box of 0.01 in redshift (left panel) or 0.5 dex in sSFR (right panel). The dashed line and dotted line indicate the ridge of the star-forming main sequence and the passive sequence, respectively.}
 \label{fig:agnz}
\end{figure*}

The frequency of LINERs ($f_{\rm LINER}$) determined within moving boxes of 0.01\,dex in redshift of central disk galaxies (light-blue line) is shown in Figure \ref{fig:agnz}. As mentioned in Section \ref{sec:fagn}, the frequency of LINERs decreases significantly with redshift owing to the selection effect. To correct this bias, we assume there is no redshift evolution of $f_{\rm LINER}$ in such a narrow redshift range of 0.02$-$0.05. Then, we use the average value of $f_{\rm LINER}$ within $0.02<z<0.025$ as the reference value and weight each galaxy by a correction factor such that the weighted $f_{\rm LINER}$ is about flat with redshift, shown as the dark-blue line in the same plot. The right panel of Figure \ref{fig:agnz} shows the $f_{\rm LINER}$ as a function of sSFR for central disk galaxies, before (light-blue line) and after (dark-blue line) this selection correction.

\end{document}